\documentclass{paper}
\usepackage{amsfonts,amsmath,latexsym,amscd,graphicx,amssymb}

\newcommand{\be}{\begin{equation}}
\newcommand{\ee}{\end{equation}}
\newcommand{\ba}{\begin{eqnarray}}
\newcommand{\ea}{\end{eqnarray}}

\newcommand{\pa}{\partial}
\newcommand{\f}{\frac}

\begin{document}

\title{Exact solutions for small-amplitude capillary-gravity water
waves}

\author{\normalsize Delia IONESCU-KRUSE\\
\normalsize Institute of Mathematics of
the Romanian Academy,\\
\normalsize P.O. Box 1-764, RO-014700, Bucharest,
 Romania\\
\normalsize E-mail: Delia.Ionescu@imar.ro\\[10pt]}

 \date{}

\maketitle
\begin{abstract}
We present explicit solutions for the ordinary differential equations
system describing the motion of the particles beneath
small-amplitude capillary-gravity  waves which propagate on the
surface of an irrotational water flow with a flat bottom. The
required computations involve elliptic integrals of first kind, the
Legendre normal form and a solvable Abel differential equation of
the second kind.

\end{abstract}

\section{Introduction}

We consider the problem of water waves in a domain of finite depth
bounded above by a free surface and under the combined effects of
gravity and surface tension. We suppose that the water flow is
irrotational. Mathematically, the problem is formulated as a free
boundary problem for incompressible Euler equations with the
irrotational condition. After rewriting the equations in an
appropriate non-dimensional form, we have two non-dimensional
parameters $\delta$ and $\epsilon$, the shallowness parameter and
the amplitude parameter, respectively, and another non-dimensional
parameter $W_e$ called Weber number, which comes from the surface
tension on the free surface. We simplify the governing equations
with a linearization which is slightly different from the classical
case in line with the Stokes condition for irrotational flows
(see, for example, \cite{cev}, \cite{cv}). By this linearization,
we obtained a parameter $c_0$ by which we can describe different
backward flows in the irrotational case:
 still water ($c_0=0$), favorable uniform current $c_0>0$, adverse
uniform current $c_0<0$.\\
 Further, we get the general solution of
the linearized problem. Notice that there are only a few explicit
solutions to the nonlinear governing equations: for gravity water
waves, Gerstner's solution\footnote{This solution was
independently re-discovered later by Rankine \cite{rankina}.
Modern detailed descriptions of this wave are given in the recent
papers \cite{c2001a} and \cite{henry4}.}\cite{gerstner} and the
edge wave solution related to it (see \cite{c2001b}), for
capillary water waves,
 Crapper's solution  \cite{crapper} and its generalization in the case of finite
 depth (see  \cite{kinn}).\\
After getting the general solution of the linearized problem we
investigate the nonlinear equations of the motion of the fluid
particles. In the case the constant $c_0$ equals the
non-dimensional speed of propagation of the linear wave, the
required computations involve elliptic integrals of first kind and
their Legendre's normal form. The exact solutions obtained in this
case contain in their expressions Jacobian elliptic functions.
Only one solution is presented in detail, the others will be
presented in a future paper. In the case the constant $c_0$ is
different from the non-dimensional speed of propagation of the
linear wave, the computations involve a solvable Abel
differential equation of the second kind.  \\
In the both cases we remark that the obtained solutions are not
closed curves. This result is in the line with  the recent results
obtained  for capillary-gravity water waves by using phase-plane
considerations for the nonlinear system describing the particle
motion (see \cite{henry2}, \cite{henry3}). By the same method see
also the results obtained for gravity water waves in \cite{cev},
\cite{cv} and for constant vorticity gravity water waves in
\cite{ehrnst}, \cite{ev}. Beside the phase-plane analysis, the
exact solutions allow a better understanding of the dynamics (see
 \cite{io}, \cite{io2}). The same type of results are obtained
for the governing equations without linearization, by analyzing a
free boundary problem for harmonic functions in a planar domain
(see \cite{c2007} for Stokes waves, \cite{CE3} for solitary waves
and \cite{henry} for deep-water Stokes waves) or by applying local
bifurcation theory (see \cite{w2} for small-amplitude waves with vorticity).\\
The existence of regular periodic travelling waves with vorticity
was recently established (see \cite{CS}, \cite{w}). For steady
periodic gravity waves the symmetry is known to be ubiquitous (see
\cite{cew}, \cite{hur}). The study of the  symmetry of rotational
water waves was initiated in the papers \cite{CE1}, \cite{CE2};
for irrotational flows see also \cite{okamoto}. However, exact
information about the flow beneath such waves, is not readily
available even in the irrotational case. This paper addresses this
issue.

\section{Small-amplitude approximation of the
  water-wave problem}

The water flow under consideration is two-dimensional, bounded by
a rigid horizontal surface below at $z=0$ and a free surface above
 at $z=h_0+\eta(x,t)$, with $h_0>0$ a constant.
 The undisturbed water surface is
  $z=h_0$. Let $(u(x,z,t),
v(x,z,t))$ be the velocity of the water and $p(x,z,t)$ be  the
pressure.
  Water can be
assumed to be inviscid fluid, even though it is slightly viscous.
 In problems of water
waves it is also reasonable to assume that the fluid is
incompressible  (constant density $\rho$) (\cite{lighthill}),
which implies the equation of mass conservation (MC). A
capillary-gravity wave is influenced by the effects of surface
tension and gravity, as well as by the fluid inertia. The surface
tension will play a role in the formulation of the boundary
conditions but not in the equations of motion valid in the fluid
domain. For the capillary-gravity water waves, the appropriate
equations of motion are Euler's equations
(EE)(\cite{johnson-carte}).  The boundary conditions for the water
wave problem are the kinematic boundary conditions as well as the
dynamic boundary condition. The kinematic boundary conditions
(KBC) express the fact that the same particles always form the
free water surface and that the fluid is assumed to be bounded
below by a hard horizontal bed $z=0$. The dynamic boundary
condition (DBC) express the fact that the difference of pressure
on the two sides of the surface $\eta$ is balanced by the effects
of surface tension.
 Thus, the boundary value problem
for capillary-gravity water waves is:
\begin{equation}
\begin{array}{c}
\begin{array}{c}
u_t+uu_x+vu_z=-\f1{\rho} p_x\\  v_t+uv_x+vv_z=-\f1{\rho} p_z-g\\
\end{array}
\quad \quad \quad \quad \textrm{ (EE) }\\
 \qquad \qquad u_x+v_z=0  \qquad \qquad \qquad \qquad \textrm{ (MC)  }\\
\begin{array}{c}
  v=\eta_t+u\eta_x \, \, \textrm{ on }\,
z=h_0+\eta(x,t)\\
 v=0 \, \,
\textrm { on } z=0
\end{array}
\quad \,\,\, \textrm{ (KBC) }
\\
\qquad
 p=p_0-\f{\Gamma}{R}, \,  \textrm{ on } z=h_0+\eta(x,t)
  \quad \quad \textrm{ (DBC)} \end{array} \label{e+bc}
\ee where  $g$ is the constant gravitational acceleration, $p_0$
is the constant atmospheric pressure, the parameter $\Gamma(>0)$
is the coefficient of surface tension and $\f{1}{R}$ is the mean
curvature (up to a factor 1/2) of the surface.
 For the surface defined as a
function $\eta(x,t)$, the mean curvature has the following
expression \be \f{1}{R}=\f{\eta_{xx}}{(1+\eta^2_x)^{3/2}} \ee
 In respect of the well-posedness for the
initial-value problem for (\ref{e+bc}) there has been significant
recent progress, see \cite{shkoller} and the references therein.\\
A key quantity in fluid dynamics is   the \emph{curl} of the
velocity field, called vorticity. For two-dimensional flows
we denote the scalar
vorticity of the flow by \be \omega(x,z)=u_z-v_x
\label{vorticity}\ee
In what follows we consider a flow
 which is uniform
 with depth, that is, described by a zero vorticity  (irrotational case).

We search for a linear approximation of the water-wave problem
(\ref{e+bc}). First the system (\ref{e+bc}) is non-dimensionalized
by  making use of the following scales: the undisturbed depth of
water $h_0$, as the vertical scale, a typical wavelength
$\lambda$, as the horizontal scale, and $\sqrt{gh_0}$ as  the
scale of the horizontal component of the velocity. The surface
wave itself leads to the introduction of a typical amplitude of
the wave $a$.
  For more details see
\cite{johnson-carte}. Thus, we define the set of non-dimensional
variables
\begin{equation}
\begin{array}{c}
x\mapsto\lambda x,  \quad z\mapsto h_0 z, \quad \eta\mapsto a\eta,
\quad t\mapsto\f\lambda{\sqrt{gh_0}}t,\\
  u\mapsto  \sqrt{gh_0}u,
\quad v\mapsto h_0\f{\sqrt{gh_0}}{\lambda}v
\end{array} \label{nondim}\end{equation}
where, to avoid new notations, we have used the same symbols for
the non-dimensional variables  $x$, $z$, $\eta$, $t$, $u$, $v$, on
the right-hand side.\\
 We set the constant water density $\rho=1$
and let us now define the non-dimensional pressure. If the water
would be stationary, that is, $u\equiv v \equiv 0$, from the
equations (EE) and (DBC) with $\eta=0$, $\Gamma=0$, we get for a
non-dimensionalised $z$, the hydrostatic pressure $p_0+ g
h_0(1-z)$. Thus, the non-dimensional pressure is defined  by
\begin{equation} p\mapsto p_0+ g h_0(1-z)+ g h_0 p
\label{p}\end{equation} Taking into account (\ref{nondim}) and
(\ref{p}) the two-dimensional capillary-gravity waves on water of
finite depth are described, in non-dimensional variables, by the
following boundary value problem
\begin{equation}
\begin{array}{cc}
u_t+uu_x+vu_z=- p_x&\\
  \delta^2(v_t+uv_x+vv_z)=- p_z&\\
 u_x+v_z=0&\\
v=\epsilon(\eta_t+u\eta_x)  \,& \textrm{ on }\,
z=1+\epsilon\eta(x,t)\\
  p=\epsilon\left[\eta-\left(\f{\Gamma}{g\lambda^2}\right)
  \f{\eta_{xx}}{(1+\epsilon^2\delta^2\eta^2_x)^{3/2}}\right] \, & \textrm{ on }\,
z=1+\epsilon\eta(x,t)\\
 v=0 \, &
\textrm { on } z=0
 \end{array}
\label{e+bc'} \end{equation}  where we have introduced the
amplitude parameter $\epsilon=\f a{h_0}$ and the shallowness
parameter $\delta=\f {h_0}{\lambda}$. \\
For irrotational flows the
 vorticity equation (\ref{vorticity}) writes in non-dimensional
 variables (\ref{nondim})
 as
 \be
 u_z=\delta^2v_x\label{vor1}
 \ee
We observe now that, on $z=1+\epsilon\eta$, both $v$ and $p$ are
proportional to $\epsilon$. Thus, with the following scaling   of
the non-dimensional variables, (avoiding again the introduction of
a new notation),
\begin{equation} p\mapsto \epsilon p,\quad
(u,v)\mapsto\epsilon(u,v) \label{scaling}\end{equation}  the
problem (\ref{e+bc'}) becomes \begin{equation}
\begin{array}{cc}
u_t+\epsilon(uu_x+vu_z)=- p_x&\\  \delta^2[v_t+\epsilon(uv_x+vv_z)]=- p_z&\\
 u_x+v_z=0&\\
  v=\eta_t+\epsilon u\eta_x  \, & \textrm{ on }\,
z=1+\epsilon\eta(x,t)\\
p=\eta-\left(\f{\Gamma}{g\lambda^2}\right)
\f{\eta_{xx}}{(1+\epsilon^2\delta^2\eta^2_x)^{3/2}}
\, & \textrm{ on }\,
z=1+\epsilon\eta(x,t)\\
 v=0 \, &
\textrm { on } z=0
 \end{array}
\label{e+bc1''} \end{equation} and the equation (\ref{vor1}) keeps
the same form.  Therefore, the system which describes the full
problem in the irrotational case is given by (\ref{e+bc1''} )+(\ref{vor1}).\\
It is conventional to write $\f{\Gamma}{\rho g
\lambda^2}=\delta^2W_e$, with $W_e=\f{\Gamma}{\rho g h_0^2}\,\,$ a
Weber number.
This  parameter is used to measure
the size of the surface tension contribution.

By letting $\epsilon\rightarrow 0$, $\delta$ and $W_e$ being
fixed, we obtain a linear approximation of
 the scaled
version (\ref{e+bc1''})+(\ref{vor1}) of our problem, that is,
\begin{equation}
\begin{array}{cc}
u_t+p_x=0&\\ \delta^2v_t+ p_z=0&\\
 u_x+v_z=0&\\
 u_z-\delta^2v_x=0&\\
v=\eta_t  \, & \textrm{ on }\,
z=1\\
  p=\eta-\delta^2 W_e\eta_{xx} \, & \textrm{ on }\,
z=1\\
 v=0 \, &
\textrm { on } z=0
\end{array}
\label{small} \end{equation}
 From the first three equations  in
(\ref{small}), we get that \be
v_{zzt}=-u_{xzt}=p_{xxz}=-\delta^2v_{xxt} \ee Therefore, \be
v_{zzt}+\delta^2 v_{xxt}=0\ee and thus \be
v_{zz}+\delta^2v_{xx}=f(x,z)\label{17} \ee where $f$ is an
arbitrary function. Taking into account the forth equation  in
(\ref{small}), we obtain that \be \delta^2 v_{xx}=u_{zx}
\label{18}\ee Introducing (\ref{18}) into (\ref{17}), we have \be
(v_z+u_x)_z=f(x,z) \ee and in view of the third equation in
(\ref{small}), we get  that \be f(x,z)=0 \ee The equation
(\ref{17}) becomes \be v_{zz}+\delta^2v_{xx}=0\label{17'}\ee We
apply the method of separation of variables, seeking a solution of
this equation in the form \be v(x,z,t)=F(x,t)G(z,t) \label{19}\ee
Substituting (\ref{19}) into the equation (\ref{17'}), we find \be
F\f{\pa^2 G}{\pa z^2}+\delta^2 G\f{\pa^2F}{\pa x^2}=0 \ee thus,
\be\f{1}{G}\f{\pa^2 G}{\pa z^2}=-\delta^2\f{1}{F}\f{\pa^2 F}{\pa
x^2}\ee We observe in the above equation that the left hand side
does not depend on $z$ and the right hand side does not depend on
$x$. Therefore, each side must be a constant, say \be
\f{1}{F}\f{\pa^2 F}{\pa x^2}=-k^2,\quad \f{1}{G}\f{\pa^2 G}{\pa
z^2}=k^2\delta^2 \label{20} \ee where $k\geq 0 $ is a constant
that might depend on time. With the above choice, the solutions of
the equations in (\ref{20}) are \ba
F(x,t)&=&A\sin(kx)+B\cos(kx)\nonumber\\
G(x,t)&=&Ce^{k\delta z}+De^{-k\delta z},\label{21} \ea where $A$,
$B$, $C$, $D$ are constants depending on time. We made this choice
of the sign of the constant in the equations (\ref{20}), in order
to obtain this wave-like solution (\ref{21}) propagating in the
$x$-direction.  On the bed $z=0$, by the last equation in
(\ref{small}), we have $v=0$, thus $C=-D$. Therefore, \be
v(x,z,t)=\sinh (k\delta
z)\left(\mathcal{A}\sin(kx)+\mathcal{B}\cos(kx)\right) \ee where
we introduced $C=-D$ into the constants $\mathcal{A}$ and
$\mathcal{B}$.\\
Taking now into the account the fifth equation in (\ref{small}),
on $z=1$ we get \be \sinh
(k\delta)\left(\mathcal{A}\sin(kx)+\mathcal{B}\cos(kx)\right)=\eta_t
\ee which implies \be
\left(\mathcal{A}\sin(kx)+\mathcal{B}\cos(kx)\right)=\f{\eta_t}{\sinh(k\delta)}\ee
Hence,  \be v(x,z,t) =\f{1}{\sinh(k\delta)}\sinh (k\delta
z)\eta_t\label{22}\ee For the component $u$ of the velocity field,
 taking into account (\ref{22}) and the
fourth equation of the system (\ref{small}), we obtain \be
u(x,z,t)=\f{\delta}{k\sinh (k\delta)}\cosh (k\delta
z)\eta_{tx}+\mathcal{F}(x,t) \label{23}\ee where
$\mathcal{F}(x,t)$ is an arbitrary function. The components $u$
and $v$ of the velocity have to fulfill also the third equation in
(\ref{small}), hence, in view of (\ref{22}) and (\ref{23}),  \be
\f{\delta}{k\sinh (k\delta)}\cosh (k\delta
z)\eta_{txx}+\f{\pa\mathcal{F}(x,t)}{\pa x}=-\f{k\delta }{\sinh
(k\delta)}\cosh (k\delta z)\eta_{t} \label{23'}\ee The above
relation must hold for all values of $x\in\mathbf{R}$, and $0\leq
z\leq 1$. It follows \be \f{\pa\mathcal{F}(x,t)}{\pa
x}=0\label{24}\ee and \be \eta_{txx}+k^2\eta_t=0 \label{25}\ee We
seek periodic travelling wave solutions, thus, for the equation
(\ref{25}) with \be k=2\pi\ee
  we choose the following solution \be
\eta(x,t)=\cos(2\pi(x-ct)) \label{26}\ee where $c$ represents the
non-dimensional speed of propagation of the linear wave and is to be determined.\\
 From (\ref{24}) the function
$\mathcal{F}(x,t)$ is independent of $x$, therefore we will denote
this function by $\mathcal{F}(t)$. \\
We return now to the systems (\ref{small}) in order to find the
the expressions of the pressure. Taking into account the first two
equations in (\ref{small}) and the expressions of the velocity
field from above, we obtain  \be p(x,z,t)=\f{2\pi\delta
c^2}{\sinh(2\pi\delta)}\cosh(2\pi\delta z)\cos(2\pi(x-ct))+
x\mathcal{F}'(t) \label{29}\ee  On the free surface $z=1$ the
pressure (\ref{29}) has to fulfill the sixth equation of the
system (\ref{small}). Hence, in view of (\ref{26}), we get \be
2\pi\delta c^2\coth(2\pi\delta)\cos(2\pi(x-ct))+
x\mathcal{F}'(t)=(1+4\pi^2\delta^2W_e)\cos(2\pi(x-ct)) \ee The
above relation must hold for all values $x\in \mathbf{R}$,
therefore, we get \be \mathcal{F}(t)=\textrm{constant}:=c_0 \ee
and we provide the non-dimensional speed of the linear wave \be
c^2=\f{\tanh(2\pi\delta)}{2\pi\delta}(1+4\pi^2\delta^2W_e)=
\f{\lambda}{2\pi h_0}\left(1+
\f{4\pi^2\Gamma}{g\lambda^2}\right)\tanh\left(\f{2\pi
h_0}{\lambda}\right)\label{c}\ee We observe thus, that the speed
of propagation of the wave varies with the wavelength $\lambda$,
with the undisturbed depth $h_0$ and with the coefficient of
surface tension $\Gamma$.

Summing up,  the solution of the linear system (\ref{small}) is
\be
 \begin{array}{llll}
 \eta(x,t)=\cos(2\pi(x-ct))\\
 p(x,z,t)=\f{2\pi\delta
c^2}{\sinh(2\pi\delta)}\cosh(2\pi\delta z)\cos(2\pi(x-ct))\\
u(x,z,t)=\f{2\pi\delta c}{\sinh (2\pi\delta)}\cosh (2\pi\delta
z)\cos(2\pi(x-ct))+c_0\\
v(x,z,t)=\f{2\pi c}{\sinh(2\pi\delta)}\sinh (2\pi\delta
z)\sin(2\pi(x-ct)) \end{array}\label{solrot0}\ee with $c$ given by
(\ref{c}).

\section{Exact solutions to the nonlinear equations of the motion of fluid particles}

Let $\left(x(t), z(t)\right)$ be the path of a particle in the
fluid domain, with location $\left(x(0), z(0)\right):=(x_0,z_0)$
at time $t=0$. Taking into account (\ref{solrot0}), the motion of
the particle is described by the following system of nonlinear differential
equations
  \be\left\{\begin{array}{ll}
 \f{dx}{dt}=u(x,z,t)=\f{2\pi\delta c}{\sinh (2\pi\delta)}\cosh (2\pi\delta
z)\cos(2\pi(x-ct))+c_0\\
 \f{dz}{dt}=v(x,z,t)=\f{2\pi c}{\sinh(2\pi\delta)}\sinh (2\pi\delta
z)\sin(2\pi(x-ct))
 \end{array}\right.\label{diff2}\ee
 The right-hand side of the differential system (\ref{diff2})
 is smooth and bounded, therefore, the unique solution of the Cauchy
 problem with initial data $(x_0,z_0)$ is defined globally in
 time.\\
 Notice that the constant $c_0$ is the average of the horizontal
fluid
 velocity over any horizontal
 segment of length 1, that is,
 \be
 c_0=\f 1 {1}\int_{x}^{x+1}u(s,z,t)ds,
 \ee
 representing therefore the strength of the underlying uniform
 current. Thus, $c_0=0$  will correspond to a region of still water with
 no underlying current,
 $c_0>0$ will characterize a favorable uniform current and  $c_0<0$
 will characterize an adverse uniform current.\\
To study the exact solution of the system (\ref{diff2}) it is
 more convenient to re-write it in the following moving frame
 \be
 X=2\pi(x-ct),\quad  Z=2\pi\delta z \label{frame}
 \ee
This transformation yields \be\left\{\begin{array}{ll}
 \f{dX}{dt}=\f{4\pi^2\delta c}{\sinh(2\pi\delta)}\cosh(Z)\cos(X)+2\pi(c_0-c)\\
 \f{dZ}{dt}=\f{4\pi^2\delta c}{\sinh(2\pi\delta)} \sinh(Z)\sin(X)
 \end{array}\right.\label{diff3}\ee
\textbf{I)} $\mathbf{c_0=c}$\\
In this case,  differentiating with respect to $t$, the system
(\ref{diff3}) can be written in the following form:
\be\left\{\begin{array}{ll}
 \f{d^2 X}{dt^2}=-\f{8\pi^4\delta^2 c^2}{\sinh^2(2\pi\delta)}\sin(2X)\\
 \f{d^2Z}{dt^2}=\f{8\pi^4\delta^2 c^2}{\sinh^2(2\pi\delta)} \sinh(2Z)
 \end{array}\right.\label{30}\ee
This system integrates to  \be \left\{\begin{array}{ll}
 \left(\f{dX}{dt}\right)^2=\f{8\pi^4\delta^2 c^2}{\sinh^2(2\pi\delta)}
 \cos(2X)+c_1\\
 \left(\f{dZ}{dt}\right)^2=\f{8\pi^4\delta^2 c^2}{\sinh^2(2\pi\delta)}
 \cosh(2Z)+c_2
 \end{array}\right.\label{31}
\ee $c_1$, $c_2$ being the integration constants.\\
For the first equation in (\ref{31}) we use the substitution
 \be \tan(X)=y\, ,\,
 \cos(2X)=\f{1-y^2}{1+y^2}\,,\,\sin
(2X)=\f{2y}{1+y^2}\,, \, dX=\f{1}{1+y^2}dy\label{substitution}\ee
 In the new variable, the first equation in (\ref{31})
 takes the form
 \be \left(\f{dy}{dt}\right)^2=\f{8\pi^4\delta^2
 c^2}{\sinh^2(2\pi\delta)}(1-y^4)+c_1(1+y^2)^2\label{32}
 \ee
We denote by \be
a^2:= \f{8\pi^4\delta^2
 c^2}{\sinh^2(2\pi\delta)}
\label{a2}\ee
 The solution of the equation (\ref{32})
involves an elliptic integral of first kind:
 \be \pm\int\f{dy}{\sqrt{(c_1-a^2)y^4+2c_1y^2+c_1+a^2}}=t\label{33}\ee The
elliptic integral of first kind from (\ref{33}) may by reduced to
Legendre's normal form. In order to do this we consider first the
substitution \be y^2=s \label{39} \ee Therefore, the left hand
side in
 (\ref{33})
becomes \be \pm\int\f{dy}{\sqrt{(c_1-a^2)y^4+2c_1y^2+c_1+a^2}}=\pm
\int\f{ds}{2\sqrt{(c_1-a^2)s(s+1)\left(s+\f{c_1+a^2}{c_1-a^2}\right)}}
\label{40}\ee Further, we  introduce a new variable $\varphi$.
 The definition of this variable depends on the sign of $c_1-a^2$ and
$c_1+a^2$. There are three possibilities: \\
$c_1-a^2>0$,\\
 $c_1-a^2<0$ and $c_1+a^2>0$, \\
 $c_1-a^2<0$ and
$c_1+a^2<0$. \\
We present below only the second case, the
investigation of the others will be presented in a future
paper.\\
If \be c_1-a^2<0 \quad \textrm{and} \quad c_1+a^2>0\ee then,  we
introduce the variable $\varphi$ by (see \cite{smirnov} Ch. VI, \S
4, page 602) \be
s=\f{a^2+c_1}{a^2-c_1}\cos^2\varphi\label{varphi'} \ee and we get
\ba &&(c_1-a^2)s(s+1)\left(s+\f{c_1+a^2}{c_1-a^2}\right)=
\f{2a^2(a^2+c_1)^2}{(a^2-c_1)^2}\sin^2\varphi\cos^2\varphi\left[
1-k_1^2\sin^2\varphi \right] \nonumber\\
&&ds=-\f{2(a^2+c_1)}{a^2-c_1}\sin\varphi\cos\varphi d\varphi
\nonumber\ea where the constant $0<k_1^2<1$ is given by \be
k_1^2=\f{a^2+c_1}{2a^2}\ee Therefore we obtain the Legendre normal
form of the integral in (\ref{33}), that is, \be \pm\f 1
{\sqrt{2}\, a}\int\f{d\varphi}{\sqrt{1-k_1^2\sin^2\varphi }}=t
\label{40'}\ee The inverse of the integral in (\ref{40'}) is the
Jacobian elliptic function  \textit{sine amplitude} (see, for
example, \cite{byrd}), an odd periodic function of order two, \be
\textrm{ sn }\left(\pm\sqrt{2}a \,t;k_1\right): =\sin\varphi\ee
 In view of the notations (\ref{39}),
(\ref{varphi'}), we get that \be y(t)=\pm \sqrt{\f
{a^2+c_1}{a^2-c_1}}\,\textrm{cn
}\left(\pm\sqrt{2}a\,t;k_1\right)=\pm \sqrt{\f
{a^2+c_1}{a^2-c_1}}\,\textrm{cn
}\left(\sqrt{2}a\,t;k_1\right)\label{sol'}\ee cn being the
Jacobian elliptic function \textit{cosine amplitude}, an even
periodic
function of order two.  \\
 For the second equation in (\ref{31})
 we use the substitution
 \be \tanh(Z)=w\, ,\quad
 \cosh(2Z)=\f{1+w^2}{1-w^2}\,, \quad
dX=\f{1}{1-w^2}dw\label{substitution'}\ee
 In the new variable, the second equation in (\ref{31}) takes the form
 \be
\left(\f{dw}{dt}\right)^2=\f{8\pi^4\delta^2
 c^2}{\sinh^2(2\pi\delta)}(1-w^4)+c_2(1-w^2)^2\label{32'}
 \ee
The solution of the equation (\ref{32'}) involves an elliptic
integral of first kind:
 \be \pm\int\f{dw}{\sqrt{(c_2-a^2)w^4-2c_2w^2+c_2+a^2}}=t\label{33'}\ee
 where $a^2$ is the constant from (\ref{a2}). The
elliptic integral of first kind from (\ref{33'}) may by reduced to
Legendre's normal form. In order to do this we consider first the
substitution \be w^2=r \label{39'} \ee The left hand side in
 (\ref{33'})
becomes \be
\pm\int\f{dw}{2\sqrt{(c_2-a^2)w^4-2c_2w^2+c_2+a^2}}=\pm
\int\f{ds}{2\sqrt{(c_2-a^2)r(r-1)(r-\f{c_2+a^2}{c_2-a^2})}}
\label{401}\ee As in the case of the  integral in (\ref{40}), we
introduce a new variable $\phi$.
 The definition of  $\phi$ depends on the sign of $c_2-a^2$ and
$c_2+a^2$. There are three possibilities:\\
 $c_2-a^2>0$,\\
$c_2-a^2<0$ and $c_2+a^2>0$,\\
 $c_2-a^2<0$ and $c_2+a^2<0$. \\
 We
present below only the last case, the investigation of the
others  will be presented in detail in a future
paper.\\
If \be c_2-a^2<0 \quad \textrm{ and }\quad c_2+a^2<0\ee then,  we
introduce the variable $\phi$ by (see \cite{smirnov} Ch. VI, \S 4,
page 602) \be r= 1+\f{2a^2}{c_2-a^2}\sin^2\phi \label{varphi1''}
 \ee
and we get \ba
&&(c_2-a^2)r(r-1)\left(r-\f{c_2+a^2}{c_2-a^2}\right)=
 \f{4a^4}{(a^2-c_2)}\sin^2\phi\cos^2
 \phi\left(1-k^2_2\sin^2\phi\right)\nonumber\\
&&ds=\f{4a^2}{c_2-a^2}\sin\phi\cos\phi d\phi\nonumber\ea where the
constant $0<k_2^2<1$ is given by \be k_2^2=\f{2a^2}{a^2-c_2}\ee
Therefore we obtain the Legendre normal form of the integral in
(\ref{33'}), that is, \be \pm
\f{1}{\sqrt{a^2-c_2}}\int\f{d\phi}{\sqrt{1-k_2^2\sin^2\phi }}=t
\label{401'''}\ee The inverse of the integral in (\ref{401'''}) is
\be \textrm{ sn }\left(\pm\sqrt{a^2-c_2}t;k_2\right): =\sin\phi\ee
In view of the notations (\ref{39'}), (\ref{varphi1''}), we get
that \be w(t)=\pm\sqrt{1-\f{2a^2}{a^2-c_2}\textrm{ sn
}^2\left(\sqrt{a^2-c_2}t;k_2\right)}\label{sol1''}\ee
 Therefore, from (\ref{substitution}) and (\ref{substitution'}),
 the solution of the system (\ref{31}) has the following
expression  \be
\begin{array}{ll}
 X(t)=\textrm{arctan }[y(t)]\\
 Z(t)=\textrm{arctanh }[w(t)]=\f1{2}\ln\f{1+w(t)}{1-w(t)}
 \end{array}\label{34}
\ee with $y(t)$ given by (\ref{sol'})  and $w(t)$ given by
(\ref{sol1''}).
 From (\ref{frame}) and  (\ref{34}), the solution of the
system (\ref{diff2}) with the constant $c_0$ equals  the speed of
propagation of the linear wave $c$, have the following
expressions:
 \be
\begin{array}{ll}
 x(t)=ct\pm\f1{2\pi}\textrm{arctan }\left[ \sqrt{\f
{a^2+c_1}{a^2-c_1}}\,\textrm{cn
}\left(\sqrt{2}a\,t;k_1\right)\right]\\
 z(t)=\pm\f 1{2\pi\delta}\textrm{arctanh }\left[\sqrt{1-\f{2a^2}{a^2-c_2}\textrm{ sn
}^2\left(\sqrt{a^2-c_2}t;k_2\right)}\right]
 \end{array}\label{solutie6}
\ee
We remark that the curve in (\ref{solutie6}) is not a closed curve. This result is in the line with  the results obtained in
\cite{c2007}, \cite{cev}, \cite{cv}, \cite{ehrnst}, \cite{ev},
\cite{henry}, \cite{henry2}, \cite{henry3}, \cite{io}, \cite{io2}.

\textbf{II)} $\mathbf{c_0\neq c}$\\
Differentiating with respect to $t$ the system (\ref{diff3}) we
get \be
 \f{d^2 X}{dt^2}+b\tan(X)\f{dX}{dt}+a^2\sin(2X)-
 b^2\tan(X)=0\label{36}\ee
where $a^2$ is the constant from (\ref{a2} ) and \be
b:=2\pi(c_0-c)\ee
Using the substitution  (\ref{substitution}),
the equation
 (\ref{36})
takes the form \be \f{d^2
 y}{dt^2}-\f{2y}{1+y^2}\left(\f{dy}{dt}\right)^2+by\f{dy}{dt}+
 2a^2y-b^2y(1+y^2)=0
\label{37}\ee
 For
(see \cite{kamke}, 6.54, page 554)\be
p(y)=\f{dy}{dt},\label{pp}\ee the equation (\ref{37}) becomes an
Abel differential equation of the second kind \be
p\f{dp}{dy}=\f{2y}{1+y^2}p^2-by p-
 2a^2y+b^2y(1+y^2) \ee
The substitution (see \cite{kamke}, 4.11, pages 26-27) \be
u(y)=p(y)E(y),\quad \textrm{ where } \quad
E(y)=\exp\left(-\int\f{2y} {1+y^2}dy \right)=\f
1{1+y^2},\label{43}\ee brings this equation to the simpler form
\be u\f{du}{dy}=-b\f{y}{1+y^2} u-
 2a^2\f{y}{(1+y^2)^2}+b^2
 \f{y}{1+y^2}\label{38}\ee
The equation (\ref{38}) with the substitution \be
\xi=\int\left(-\f{by}{1+y^2}\right)dy=-\f b{2}\ln(1+y^2)
\label{xi}\ee can be written in the canonical form: \be
u\f{du}{d\xi}-u=
 \f{2a^2}{b}\exp\left(\f{2\xi}{b}\right)
 -b\label{normalform}\ee
The equation (\ref{normalform}) is solvable (see \cite{zaitsev},
8.,  page 111), its solution can be written out in the following
parametric form \be
\begin{array}{ll}
u(\tau)=\tau\f{C-b\ln|\tau+\sqrt{\tau^2-2a^2}|}{\sqrt{\tau^2-2a^2}}+b\\
\xi(\tau)=-b\ln\Big|\f{\sqrt{\tau^2-a^2}}{C-b\ln|\tau+\sqrt{\tau^2-2a^2}|}
\Big|\end{array}\label{44}\ee $C$ being a constant. From
(\ref{xi}) and (\ref{44}) we get the expression of $y$\be
y(\tau)=\pm\sqrt{\f{\tau^2-2a^2}{\left(
C-b\ln|\tau+\sqrt{\tau^2-2a^2}|\right)^2}-1}, \label{ytau}\ee By
(\ref{pp}) and (\ref{43}),
 we have $\f{dt}{d\tau}=\f
1{(1+y^2)u}\left(\f{dy}{d\tau}\right)$ and therefore, from
(\ref{44}) and (\ref{ytau}),  the relation between $t$ and $\tau$
is the following: \be t=\int
\f{1}{\sqrt{\tau^2-2a^2}\sqrt{\tau^2-2a^2-(C-b\ln|\tau+\sqrt{\tau^2-2a^2}|)^2}}\,d\tau
\label{t}\ee
 Thus, taking into account (\ref{substitution}), we obtain
\be X(t)=\textrm{arctan }[y(t)],\label{41}\ee with $y(\tau)$ given by (\ref{ytau})
and $\tau$ given implicitly by (\ref{t}).\\
 In
order to determine $Z(t)$ from the system (\ref{diff3}), with
(\ref{41}) in view, we write the second equation of this system in
the form \be \f{dZ}{\sinh(Z)}=\f{4\pi^2\delta
c}{\sinh(2\pi\delta)}\sin(\textrm{arctan }[y(t)])\,dt
=\f{4\pi^2\delta
c}{\sinh(2\pi\delta)}\f{y(t)}{\sqrt{1+y^2(t)}}\,dt\ee Integrating,
we get \be \ln\left[\tanh\left(\f Z{2}\right)\right]=\int
\f{4\pi^2\delta c}{\sinh(2\pi\delta)}\f{y(t)}{\sqrt{1+y^2(t)}}\,dt
+\textrm{const}\ee If \be \int \f{4\pi^2\delta
c}{\sinh(2\pi\delta)}\f{y(t)}{\sqrt{1+y^2(t)}}\,dt+\textrm{const}<
0 \label{int}\ee then \be Z(t)=2\textrm{arctanh
}\left[\exp\left(\int \f{4\pi^2\delta
c}{\sinh(2\pi\delta)}\f{y(t)}{\sqrt{1+y^2(t)}}\,dt+\textrm{const}\right)\right]
\label{42}\ee From (\ref{frame}), (\ref{41}) and (\ref{42}), the
solution of the system (\ref{diff2}) is written now as
\be\begin{array}{ll}
 x(t)=ct+\f1{2\pi}\textrm{arctan }\left[y(t)\right]\\
 z(t)=\f 1{\pi\delta}\textrm{arctanh }\left[\exp\left(\int
\f{4\pi^2\delta
c}{\sinh(2\pi\delta)}\f{y(t)}{\sqrt{1+y^2(t)}}\,dt+\textrm{const}\right)\right]
\end{array}\label{solutie''} \ee with $y(\tau)$ given by (\ref{ytau})
and $\tau$ given implicitly by (\ref{t}).
 \\
We remark that the curve in (\ref{solutie''}) is not a closed
curve. This result is in the line with  the results obtained in
\cite{c2007}, \cite{cev}, \cite{cv}, \cite{ehrnst}, \cite{ev},
\cite{henry}, \cite{henry2}, \cite{henry3}, \cite{io}, \cite{io2}.

\medskip

\medskip

\end{document}